\documentclass[aps,floatfix,twocolumn,showkeys,preprintnumbers,amsmath,amssymb]{revtex4}

\usepackage{graphicx}
\usepackage{dcolumn}
\usepackage{bm}

\begin{document}

\def\Si{\rm Si}
\def\Al{\rm Al}
\def\Mg{\rm Mg}
\def\O{\rm O}
\def\N{\rm N}
\def\S{\rm S}
\def\sialon{$\beta$-SiAlON}
\def\bSin
{$\beta$-Si$_3$N$_4$}
\def\Sin
{Si$_3$N$_4$}
\def\bsialon{$\beta$-SiAlON}
\def\Sialon
{Si$_{6-x}$Al$_x$O$_x$N$_{8-x}$}
\def\Simgoon
{Si$_{6-x}$Mg$_x$O$_{2x}$N$_{8-2x}$}
\def\Simgon
{Si$_{6-x}$Mg$_x$O$_x$N$_{8-x}$}
\def\Simgons
{Si$_{6-x}$Mg$_x$O$_x$N$_{8-2x}$S$_x$}
\def\Simgmon
{Si$_{6-x}$Mg$_{x-y}$M$_y$O$_x$N$_{8-x}$}
\def\Simgalon
{Si$_{6-x}$Mg$_{x/2}$Al$_{x/2}$O$_x$N$_{8-x}$}
\def\ecoh
{$E_{coh}$}
\def\gap{$\Delta E_g$}
\def\nef{$N(E_F)$}
\def\nefi#1{$N_{\rm #1}(E_F)$}
\def\ef{$E_F$}


\title{FP-LMTO studies of hypothetical compounds with the
$\beta$-SiAlON-like structure in Si--(Mg,Al)--O--N systems}

\author{S.V. Okatov}
\email[]{okatov@ihim.uran.ru}
\affiliation{Institute of Solid State Chemistry, 
Ural Branch of the Russian Academy of Sciences, Pervomayskaya st.,
GSP-145, 620219, Ekaterinburg, Russia}

\date{\today}

\begin{abstract}

The electronic and energy properties of $\beta$-Si$_3$N$_4$ (a),
Si$_{6-x}$Al$_x$O$_x$N$_{8-x}$ (b) and the hypothetical ordered solid
solutions Si$_{6-x}$Mg$_x$O$_{2x}$N$_{8-2x}$ (c),
Si$_{6-x}$Mg$_x$O$_x$N$_{8-2x}$S$_x$ (d), Si$_{6-x}$Mg$_x$O$_x$N$_{8-x}$
(e), Si$_{6-x}$Mg$_{x/2}$Al$_{x/2}$O$_x$N$_{8-x}$ (f) are considered by
the {\it ab-initio} band FP-LMTO method. The calculations show that the
stability of the systems decreases in the order: (a) $>$ (b) $>$ (f)
$>$ (e) $>$ (c) $>$ (d). It is established that \Simgon\ and \Simgalon\
possess non-zero values of the density of states at the Fermi energy,
which consists mainly of localized O$2p$, N$2p$ states, and the
conductivity in those solid solutions is unlikely. It is found that the
energy of the O$2s,p$ states tends to shift depending on the
coordination environment of oxygen atoms in the considered systems. The
energy of ordering of Al and Mg atoms in \Simgalon\ is estimated to be
3.15 eV/56-atomic supercell.

\end{abstract}



\maketitle

\section{Introduction}

\sialon s are solid solutions (SS) of variable composition formed from
\bSin\ by $\Al\rightarrow\Si$, $\O\rightarrow\N$ substitutions
(composition \Sialon, $x=0-4.2$). They possess a unique set of high
thermal and chemical stability, hardness, thermal conductivity,
electroinsulating characteristics, which define their application in
ceramic industry \cite{ceram}. Some peculiarities of the electronic
structure of these disordered SS are considered in papers
\cite{siching,tanaka,tanaka1}.

\begin{figure*}[t]
\begin{tabular}{cc}
\resizebox*{\columnwidth}{!}{\rotatebox{180}{\includegraphics{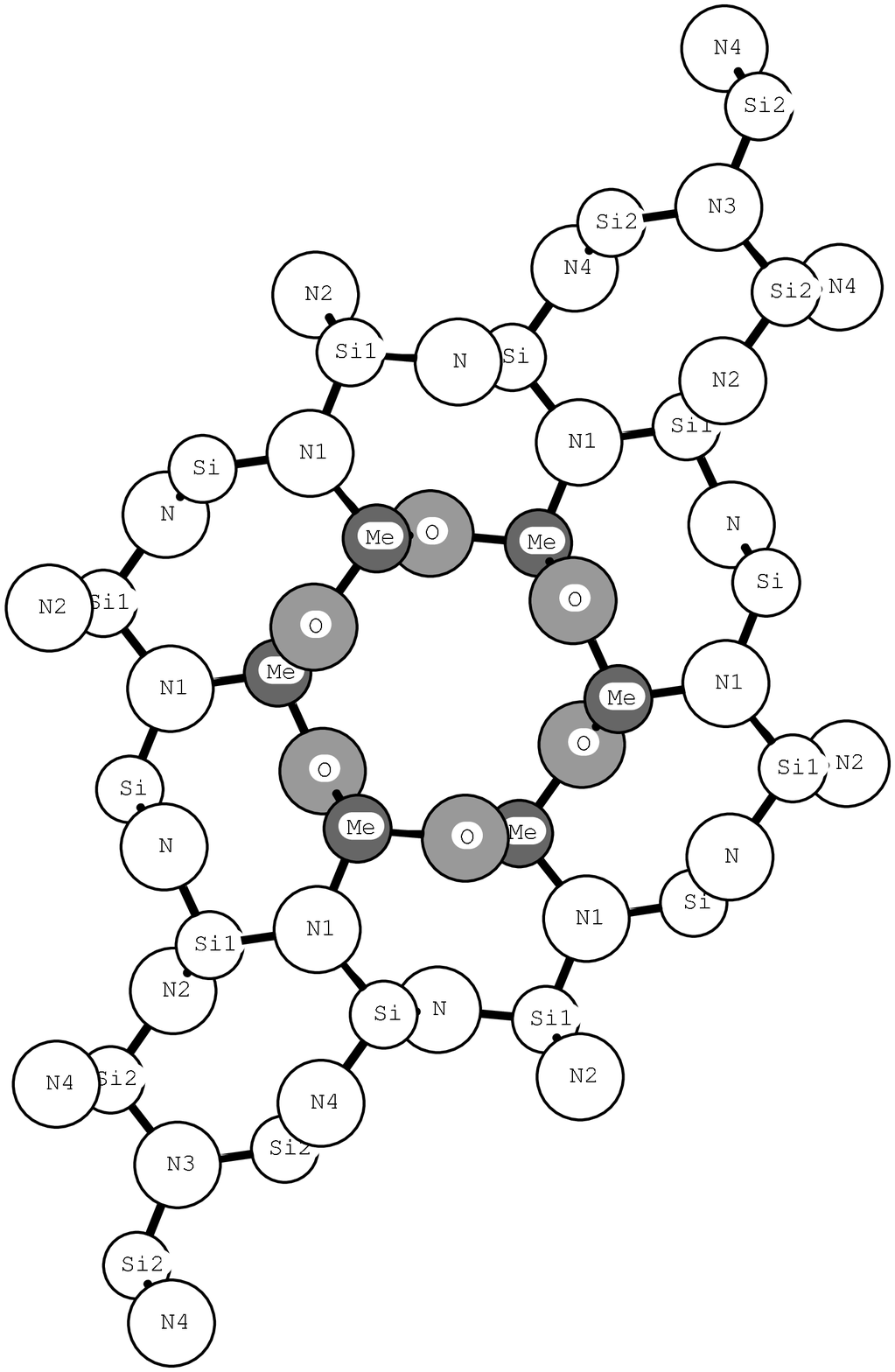}}}&
\resizebox*{\columnwidth}{!}{\rotatebox{180}{\includegraphics{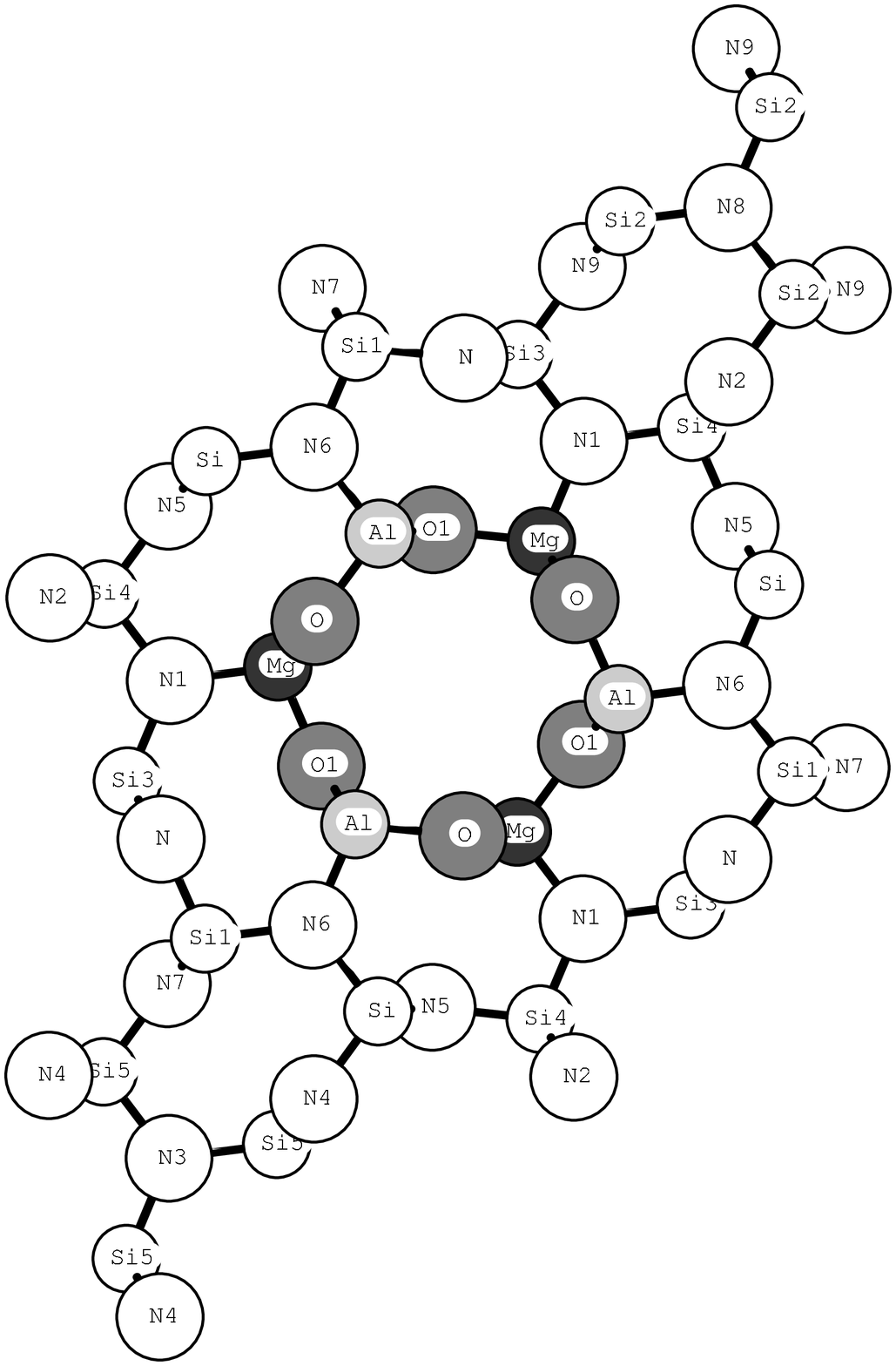}}}\\
{\LARGE a}&{\LARGE b}\\
\end{tabular}
\caption{
Models of the ordered $\Si_{6-x}{\rm Me}_x\O_x\N_{8-x}$ (a, ${\rm
Me}=\Al, \Mg$) and \Simgalon\ (b) structures containing impurity
channels. Si--Si5, N--N9 --- nonequivalent positions of Si and N,
respectively.}
\label{f.struct}
\end{figure*}

The investigation of atomic ordering effects (AOE) \cite{inmat} by the
semi-empirical tight-binding band method (EHT) allowed to predict the
existence of the so-called ``impurity channels'' in \sialon\ ---
extended along $c$ axis quasi-one-dimensional (1D) structures
constituted by 12-atomic alumoxide rings (fig. \ref{f.struct}). The
simulation of the ordered and disordered states of \sialon\ by the
cluster discrete variation method (DVM) \cite{ryz} and the band
full-potential linear muffin-tin orbitals method (FP-LMTO) \cite{flmto}
confirmed higher stability of the ordered state and revealed differences
in their electronic structure and bond indices.

The formation of 1D-structures in \sialon\ was explained in \cite{inmat}
by the ``trend'' of the system to reduce the number of Si--O, Al--O
bonds containing unfavorable antibonding states. This resulted in
localization of those states in impurity channels. Taking this fact into
account, it was supposed in \cite{dis} that the reduction of valence
electrons concentration (VEC) in the impurity channels by the partial
substitution $\Al\rightarrow{\rm M^{I,II}}$ (${\rm M^{I,II}}$ --- I, II
group elements, composition $\Si_{6-x}\Al_{x-y}{\rm
M^{I,II}}_y\O_x\N_{8-x}$) might increase the stability of the material
and induce the conductivity, which would be localized in 1D-structures.

Further investigations of AOE in the Si--Mg--O--N system by the EHT
method \cite{simgon} showed that SS of the \Simgoon\ composition could
not be formed because of unlimited growth of clusters containing Mg and O
atoms. That result was supported by experiments \cite{exp1,exp2}.
According to \cite{simgon}, the stabilization of SS in the Si--Mg--O--N
system was possible at the stoichiometric compositions \Simgon\ or
with stabilizing additives, for example Zr ($\Si_{6-x}\Mg_{x/2}{\rm
Zr}_{x/2}\O_x\N_{8-x}$) or S (\Simgons). For the latter system,
antibonding impurity states were found in the bandgap interval of \Sin,
which meant a low stability of such SS. Analogous data were obtained
for the Si--Be--O--N system \cite{sibeon}.

In this paper, the electronic and energy properties of \bSin, \Sialon,
as well as potentially stable structures \Simgoon, \Simgons, \Simgon,
\Simgalon\ obtained in \cite{simgon}, are simulated by the {\it
ab-initio} FP-LMTO method.

\section{Methods and models}

The basic \bSin\ has a hexagonal structure with the $P6_3$ space group,
lattice constants $a=7.586$, $c=2.902$\AA\ and contains 14 atoms in a
unit cell \cite{struct}. In the present calculations, we used 56-atomic
supercells ($2\times 2\times 1$) similar to those applied in
\cite{flmto}. The following compositions of the supercells are
considered: $\Si_{24}\N_{32}$, $\Si_{18}\Al_6\O_6\N_{26}$,
$\Si_{18}\Mg_6\O_{12}\N_{20}$,  $\Si_{18}\Mg_6\O_6\N_{20}\S_6$,
$\Si_{18}\Mg_6\O_6\N_{26}$,  $\Si_{18}\Mg_3\Al_3\O_6\N_{26}$, which
correspond to the stoichiometric compositions \Sin, \Sialon, \Simgoon,
\Simgons, \Simgon, \Simgalon\ ($x=1.5$, fig. \ref{f.struct}),
respectively. The atoms distribution in the cells complies with
\cite{inmat,simgon}, the lattice relaxation is not taken into account.
For the \Simgoon\ and \Simgons\ SS, N1 atoms are replaced respectively
by O and S atoms (fig. \ref{f.struct}a), which are in ``excess'' in
comparison with \Sialon, see \cite{simgon}.

The calculations are performed in the framework of the local density
approximation (LDA) by the FP-LMTO method \cite{lmfp}. The realization
of this method differs from that in \cite{flmto} by using smoothed
Hankel functions instead of unsmoothed ones, and by the representation
of the charge density. In \cite{flmto}, it was defined as $n=n_{\rm
MT}+n_{rem}$, where $n_{\rm MT}$, $n_{rem}$ --- densities within and
outside the spheres, respectively, while in \cite{lmfp} it is expressed
as $n=n_0+n_1-n_2$, where $n_0$--$n_2$ are respectively a smooth density
carried on a uniform mesh, the true density in a $Y_L$ expansion inside
each augmentation sphere, and a one-center expansion of the smooth
density inside each augmentation sphere.

The correlation energy in this work is calculated with the Barth-Hedin
formula \cite{corr}. As $a/c,b/c \sim 0.2 \ll 1$ for the chosen
supercells, the $8\times 8\times 40$ and $54\times 54\times 9$ meshes
are used for the $k$-space and charge density, respectively. To satisfy
the method's requirement of close-packing, 48 ``empty'' spheres are
included. The following values of sphere radii are chosen: 1.75 (Si, Al,
Mg), 1.47 (N, O, S) and 1.51--2.05 a.u. (for empty spheres). The atomic
orbital basis set includes $s,p,d$ states for Mg, Al, Si; $s,p$
electrons for N, O, S; and $s$ states for empty spheres.

\section{Results and discussion}

\subsection
{\bSin\ and \Sialon}

The electronic properties of \Sin\ and \Sialon\ are considered in
\cite{ryz,flmto,siching,sin1,sin2,sin3} in detail. Here we discuss only
the differences in the electronic and energy characteristics of the
ordered systems, which are due to the method
\cite{lmfp} applied.

\begin{figure*}[t]
\resizebox*{\textwidth}{!}{\rotatebox{0}{\includegraphics{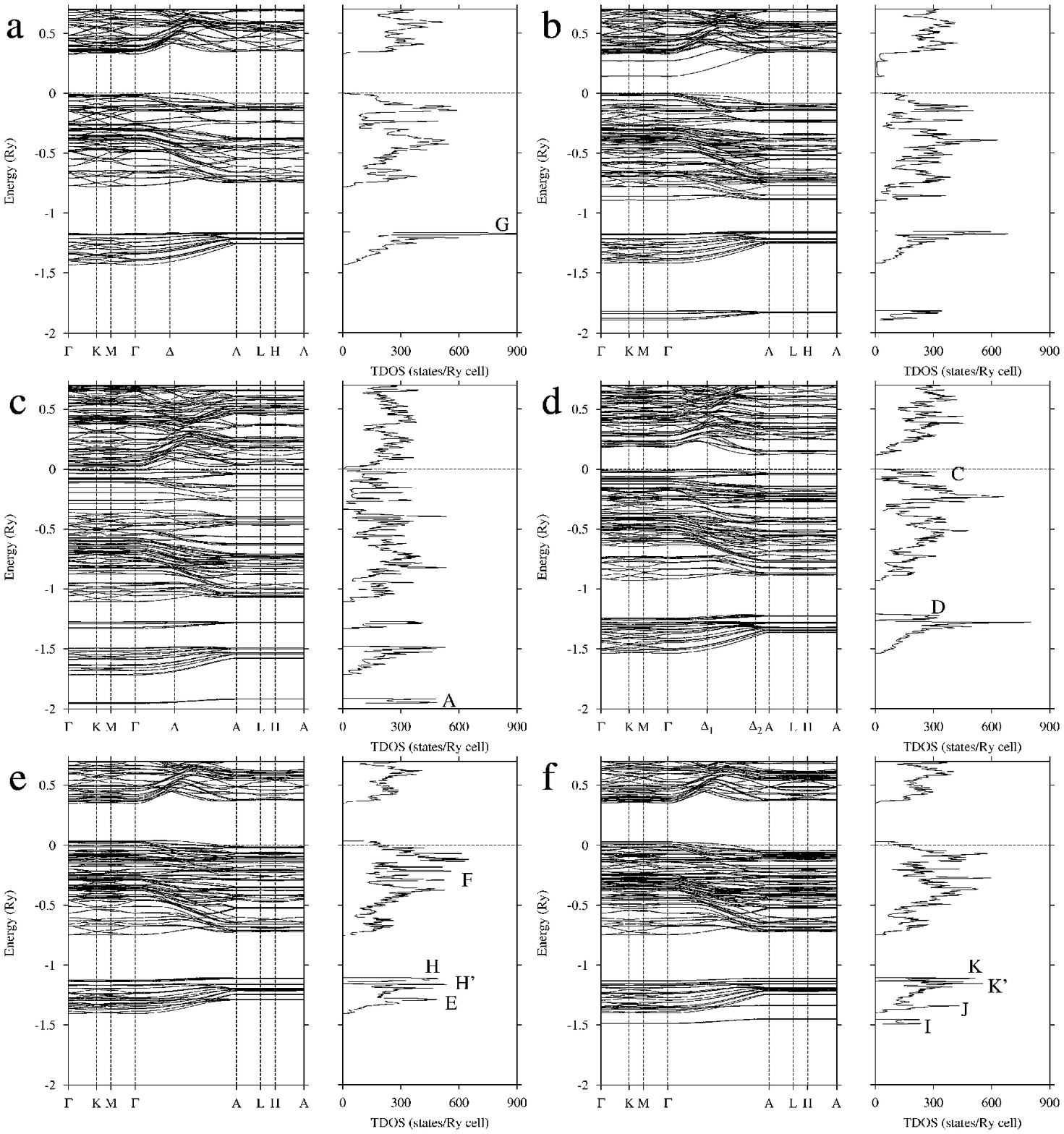}}}
\caption{
The band structure ({\it left panels}) and the total densities of
states ({\it right}) of \bSin\ (a), \Sialon\ (b) and hypothetical ordered
solid solutions \Simgoon\ (c), \Simgons\ (d), \Simgon\ (e), \Simgalon\
(f).}
\label{f.bs}
\end{figure*}

The band structures (BS) and the total densities of states (TDOS) of
\bSin\ and \Sialon\ are depicted in fig. \ref{f.bs}a,b. Their band
parameters and cohesion energies (\ecoh) are presented in table
\ref{t.sin} in comparison with the data of other publications. In good
agreement with \cite{flmto,ryz,sin1,sin2,sin3}, the valence band (VB) of
\bSin\ contains two subbands (SB) --- a low SB consisting of N$2s$ and
Si$3s,p,d$ states and a high SB, which includes N$2p$ and Si$3s,p,d$
states and separated from the conduction band (CB) by a wide bandgap
($\Delta E_g = 4.4$ eV). It is seen that, according to various authors
\cite{flmto,ryz,sin1,sin2,sin3}, the bandgap width of \Sin\ ranges from
4.1 to 5.18 eV, thus, our result agrees well with the above findings.
This value is smaller than the experimental one, but it is a
characteristic feature of the LMTO methods. A similar degree of
agreement can be found for other band parameters (see table \ref{t.sin})
except for a wider forbidden states interval (FSI). The positions,
intensities and compositions of the main peaks in this calculation are
also in agreement corresponding to that of the above parameters. Our
\ecoh\ value (5.18) also compies well with that obtained in \cite{sin1}
(5.31 eV/atom), but differs essentially from that in \cite{flmto}. A
value closer to the latter result is obtained if we choose a thin
$30\times 30\times 6$ direct space mesh. Our calculations of \bSin\ with
a 14-atomic cell and a denser mesh yield $E_{coh}=5.27$ eV/atom.

\begin{table}[t]
\caption{Bandgaps, valence band subbands (SB), forbidden states
intervals (FSI) widths  (eV, fig. \ref{f.bs}) and cohesion energies
(eV/atom) of \bSin\ and \Sialon\ obtained in these calculations and
taken from other publications.}
\centering
\begin{tabular*}{\columnwidth}{@{\extracolsep{\fill}}l|ccccccc}
\hline
Parameter&\parbox[c]{1cm}{these\\ results}&\cite{flmto}\footnote{ The
methods used: \cite{flmto} --- FP-LMTO, \cite{ryz} --- DVM, \cite{sin1}
--- pseudopotential, \cite{sin2} --- OLCAO, \cite{sin3} ---
LMTO-ASA}&\cite{ryz}\footnote{The presented parameters correspond to \Sialon\ with $x=1$.}&
\cite{sin1}&\cite{sin2}&\cite{sin3}&\parbox[c]{1cm}{Exp.\\ \cite{exp3,sine2,sine3,xray}}\\
\hline
{\bf\Sin}&&&&&&&\\
\gap&4.4&4.1&5.1&4.2&4.96&5.18&4.6-5.5\\
SB (high)&10.7&10.0&$\sim$11&10.1&9.79&8.91&---\\
FSI&5.1&4.3&$\sim$3&3.9&---&4.19&---\\
SB (low)&3.7&4.0&$\sim$5&4.2&4.12&3.49&---\\
\ecoh&5.18&6.71&---&5.31&---&---&5.93\\
\hline
{\bf \Sialon}&&&&&&&\\
\gap&1.9&2.2&4.0&\\
SB (high)&12.2&9.9&$\sim$11\\
FSI (high)&3.4&4.1&$\sim$3\\
SB (middle)&3.7&4.0&$\sim$5\\
FSI (low)&5.4&2.5&$\sim$1\\
SB (low)&1.1&1.4&$\sim$2\\
\ecoh&4.43&6.23&---\\
\hline
\end{tabular*}
\label{t.sin}
\end{table}

\bsialon\ contains three SB in the VB --- high (O$2s$, N$2s$,
Si$3s,p,d$, Al$3s,p,d$), middle (N$2s$, Si$3s,p,d$ and a small amount of
Al$3s,p,d$) and low (O$2s$, Al$3s,p,d$) --- separated from each other by
two forbidden states intervals (FSI), fig. \ref{f.bs}b. Its bandgap
($\Delta E_g=1.9$ eV), SB, FSI widths and the cohesion energy
($E_{coh}=4.43$ eV/atom) (table \ref{t.sin}) agree with the data of
other pulications \cite{flmto,ryz} as they did for \Sin\ above.
 
Generally, the present results are in good agreement with other
calculations and experiments.

\subsection{\Simgoon}

According to \cite{simgon}, the structure of the \Simgoon\ SS used in
this calculation (fig. \ref{f.struct}) is not the only possible one.
This configuration is chosen because its symmetry coincides with that of
\Sialon, which allows to consider only the chemical composition altering
effects without the symmetry correction.

The band structure and TDOS of the \Simgoon\ SS are shown in fig.
\ref{f.bs}c and some band and energy parameters are listed in table
\ref{t.oth}. One can see that the system contains Mg$3s,p,d$,
Si$3s,p,d$, N$2p$, O$2p$ states in the bandgap interval of \Sin\ and is
a semimetal with $\Delta-\Gamma$ transition. The presence of those
states causes a shift of the spectrum of the system by $\sim 4$ eV
downwards as compared with \Sin. The TDOS of \Simgoon\ contains also
two quasi-core SB of O$2s$ states. The lower one (A, fig. \ref{f.bs})
looks like that in \bsialon, whereas the higher SB (B) is specific for
this SS.

\begin{table}[t]
\caption{
The bandgap widths (\gap), population at the Fermi energy (\nef) and
cohesion energies (\ecoh) of the systems considered.}
\centering
\begin{tabular*}{\columnwidth}{@{\extracolsep{\fill}}l|ccc}
\hline
Composition&\parbox[c]{1cm}{\gap, eV}&\parbox[c]{2.2cm}{
$N(E_F)\cdot 10^{-3}$, states/eV\;cell}&
\parbox[c]{1.5cm}{\ecoh, eV/atom}\\
\hline
\Sin&4.4&0&5.18\\
\Sialon&1.9&0&4.43\\
\Simgoon&0&0&3.20\\
\Simgons&1.6&0&3.00\\
\Simgon&0&3.0&3.80\\
\Simgalon&0&2.0&4.17\\
\hline
\end{tabular*}
\label{t.oth}
\end{table}

The cohesion energy of the \Simgoon\ SS is 3.2 eV/atom, which is much
less than that for \bsialon\ and especially for \bSin. This fact
confirms the conclusion \cite{simgon} about instability of the \Simgoon\
SS.

\subsection{\Simgons}

According to \cite{simgon}, the presence of sulfur in \Simgons\ leads to
appearance of filled states in the bandgap interval, which includes
antibonding states of Si--N, Si--S bonds. Let us compare that model
with {\it ab-initio} calculations results.

As is seen from (fig. \ref{f.bs}d), the TDOS of \Simgons\ contains the
impurity states separated into a subband C composed by Si$3s,p,d$,
S$2p$, N$2p$ states and admixed to the lower edge of the CB. This
results in almost tree-times reduction of the bandgap width ($\Delta
E_g=1.6$ eV, table \ref{t.oth}) with respect to \bSin. The bandgap in
\Simgons\ is formed by the indirect transition $\Delta_1-\Delta_2$.
Another subband D of O$2s$ states appears near the N$2s$ subband. The
S$2s$ and S$2p$ states are admixed to the N$2s$ and N$2p$ SB.

In agreement with \cite{simgon}, the cohesion energy of \Simgons\ is 3.0
eV/atom, which indicates a low stability of such SS.

\subsection{\Simgon}

\begin{figure}[t]
\resizebox*{\columnwidth}{!}{\rotatebox{0}{\includegraphics{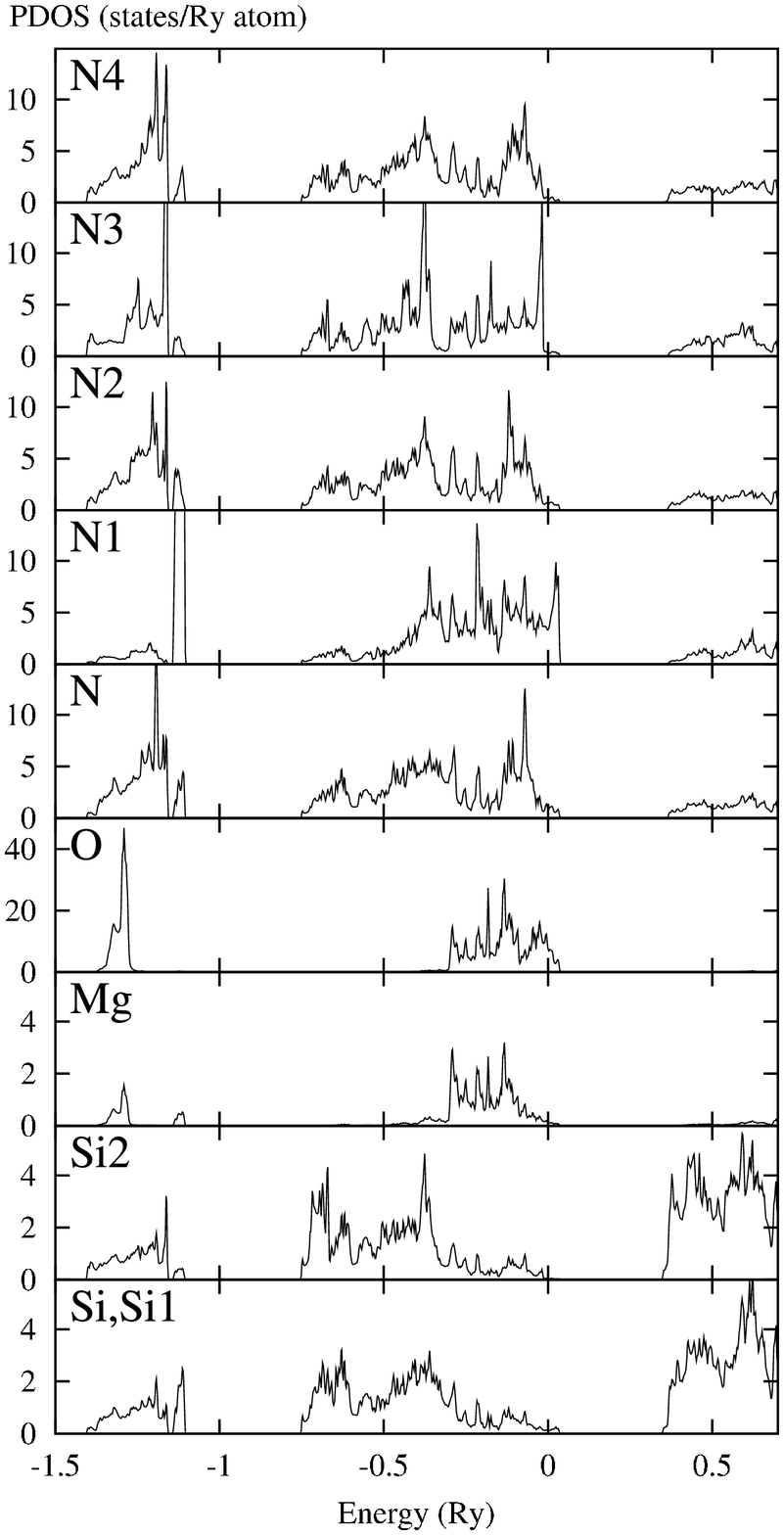}}}
\caption{The partial densities of states of \Simgon. Si--Si2, N--N4 ---
nonequivalent positions of Si, N atoms (fig. \ref{f.struct}a).}
\label{f.pdos}
\end{figure}

The band structure and TDOS of \Simgon\ are depicted in fig.
\ref{f.bs}e. The partial densities of states (PDOS) are also shown in
fig. \ref{f.pdos}. It is seen that the TDOS of \Simgon\ contains two
subbands in the VB --- high and low --- with contributions from
Si$3s,p,d$, N$2p$, Mg$3s,p,d$, O$2p$ and Si$3s,p,d$, N$2s$, Mg$3s,p,d$,
O$2s$ states respectively. The substitution $\Mg\rightarrow\Al$ in
\Sialon\ leads to electron deficiency in the resulting \Simgon. Thus,
this brings about a shift of the Fermi energy (\ef) to a lower value
and gives rise to electronic states at \ef, the number of which is
$N(E_F)=3.0\cdot 10^3$ states/eV cell (table \ref{t.oth}).

The lower edge of the CB of \Simgon\ is formed mainly by O$2p$ and to a
lesser degree by N$2p$ states. Their contributions to the density of
states at the Fermi energy (\nefi{i}, i$=$Si--Si3, N--N4, Mg, O) are 
101 (O), 46 (N1), 12 (N), 8 (N2), 5 (N4), 4 (N3) states/eV atom in the
order of distances from the impurity channel (fig. \ref{f.struct}).
Other atoms do not give any substantial number of states to \nef: 1.4
(Mg, Si, Si1), 0.5 (Si2) states/eV atom. This is indicative of a
localized character of free O- and N-states, hence, we do not expect
electroconductivity in \Simgon. A decrease in the atoms contribution
to \nef\ with the distance from the impurity channel demonstrates the
charge screening effect.

As is seen from PDOS (fig. \ref{f.pdos}), Mg- and O-states are
concentrated mainly in $[-1.4: -1.2]$ (O$2s$, Mg$3s$) and $[-0.4: 0.05]$
Ry (O$2p$, Mg$3s,p,d$) intervals (respectively, $[-19.0:-16.3]$ and
$[-5.4:0.7]$ eV). The lower Mg--O hybridized states are responsible for
formation of the E (fig. \ref{f.bs}e) peak, whereas the higher states
contribute into the F region. The degenerated Si$3p$-N$2s$ states of the
highest peak (G, fig. \ref{f.bs}a) of the lower subband in \Sin\ are
split into H and H' peaks (fig. \ref{f.bs}e) in \Simgon. This is due
to the charge nonequivalence of Si--Si2 and N--N4 atoms (fig.
\ref{f.struct}). The greatest contribution to the H peak is made by N1
atoms; it decreases with the distance from the impurity channel (fig.
\ref{f.pdos}). On the contrary, the contribution to the H' peak
increases.

The cohesion energy calculation of \Simgon\  (3.8 eV/atom) revealed a
higher stability of this SS in comparison with the \Simgoon\ and
\Simgons\ systems (table \ref{t.oth}).

\subsection{SS of the general composition \Simgmon}

\begin{figure}[t]
\resizebox*{\columnwidth}{!}{\rotatebox{270}{\includegraphics{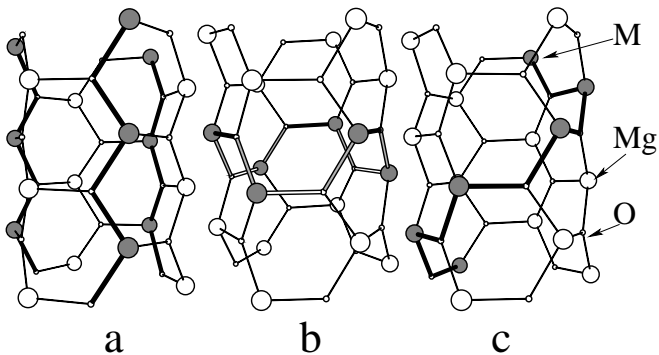}}}
\caption{Possible distributions of Mg and M atoms in \Simgmon\ within
the impurity channel: quasi-linear (a), ring-like (b) and spiral (c).}
\label{f.sol}
\end{figure}

As was noted above, O, N atoms have a large amount of states at \ef\ in
the \Simgon\ SS, but the conductivity is unlikely in this SS mainly
because of a low contribution of Mg atoms to \nef. It may be supposed
that the conductivity can be increased by partial substitution
Mg$\rightarrow$M (composition \Simgmon), which favors the formation of
electronic states at \ef. In this case the conductivity along
--M--O--M--O-- chains can be expected. This means that, depending on the
stacking type of M and Mg atoms, quasi-linear, ring-like or spiral
conductors can be formed at $y=x/3-x/2$, which are separated from each
other by insulating --Mg--O--Mg--O-- chains, fig. \ref{f.sol}.

The most interesting case is that of spiral structures, which may be
interpreted as nano-sized solenoids. A potential difference applied to
that solenoid may induce the magnetic field localized within the
impurity channel and directed along the $c$ axis. This effect can be
used, in particular, for producing electronic memory. For instance, by
placing the magnetic atoms into the cavity of the impurity channel and
altering the juice direction, it may be possible to change the direction
of their magnetic moments vectors. This may be interpreted as the
changing of the memory element state ($0\rightarrow1$ or
$1\rightarrow0$).

At present, the search for the compositions in the \Simgmon\ systems is
in progress.The \Simgalon\ SS is considered below as an example.

\subsection{\Simgalon}

In this paper we study the \Simgalon\ SS with the quasi-linear type of
ordering of Al and Mg atoms (fig. \ref{f.struct}b, \ref{f.sol}a).

The substitution of Al atoms for Mg in \Simgon\ causes the following
changes in the electronic and energy properties (fig. \ref{f.bs}f, table
\ref{t.oth}) of the resulting \Simgalon. Subband I of quasi-core
O$2s$ states is formed, which is similar that in \Sialon\ one, but the
first SB lies in a higher energy interval than the second one (fig.
\ref{f.bs}b,f). Pick J shifts by $\sim 0.7$ eV downword with respect to
peak E in \Simgon, and the distance between peaks K and K' shortens in
comparison with H and H' (fig. \ref{f.bs}e,f). The TDOS at
\ef\ decreases ($2.0\cdot 10^3$ states/eV cell) in \Simgalon\ with
respect to \Simgon\ (table \ref{t.oth}), which is due to partial filling
of empty O$2p$ states in \Simgon\ occurring when Al atoms replace Mg. In
general, the electronic structure of \Simgalon\ has an intermediate
shape between \Sialon\ and \Simgon, but the splitting of O$2s,p$ states,
which occurs in \bsialon\ when high symmetry structures (impurity
channels) are formed, is not observed.

\begin{figure*}[p]
\resizebox*{\textwidth}{!}{\rotatebox{0}{\includegraphics{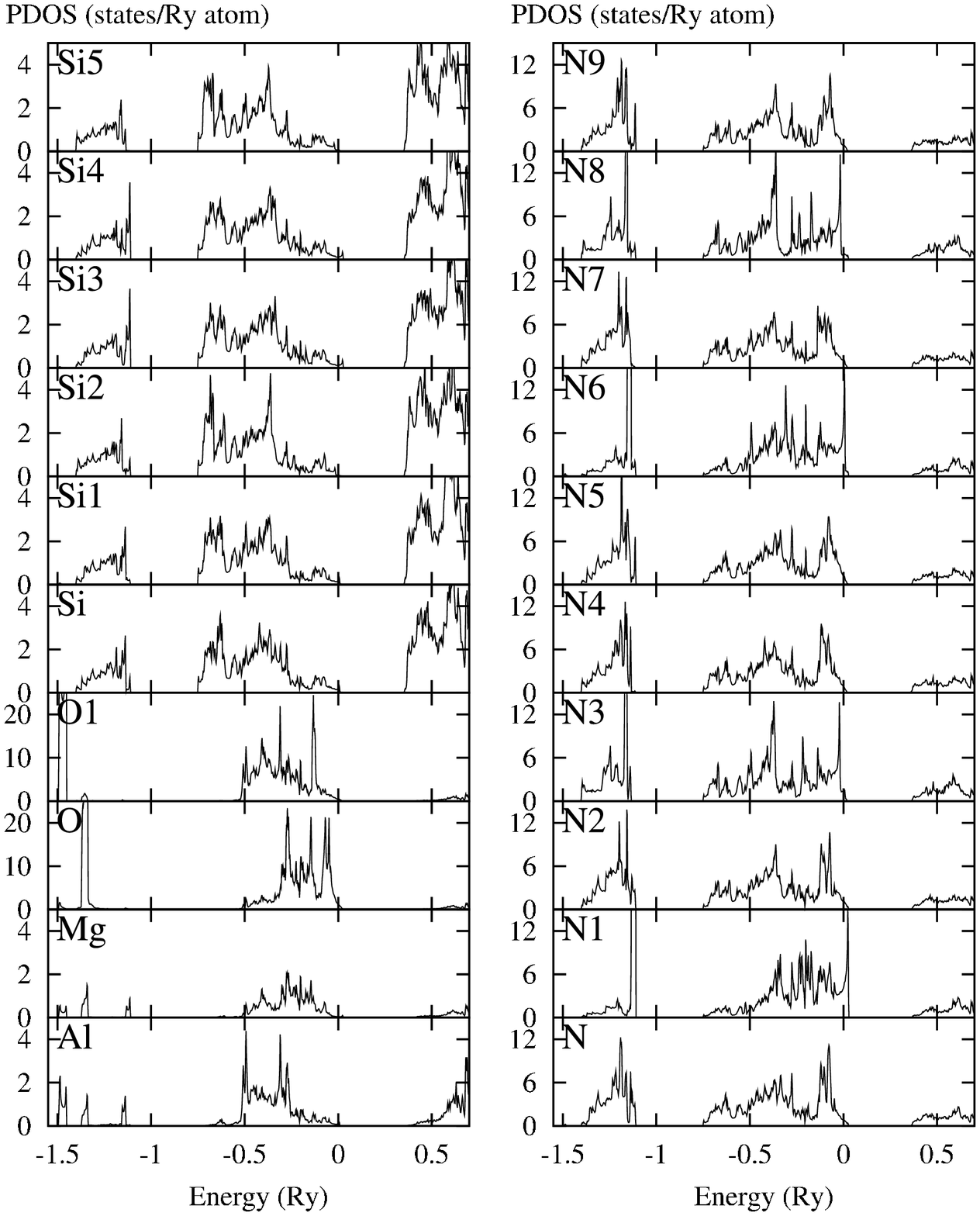}}}
\caption{The partial densities of states of \Simgalon. Si--Si5, N--N9
--- nonequivalent positions of atoms Si, N (fig. \ref{f.struct}b).}
\label{f.alp}
\end{figure*}

The analysis of PDOS (fig. \ref{f.alp}) shows that peaks I, J (fig.
\ref{f.bs}) are constituted by O$2s$ states of O1 and O atoms
respectively. Taking into account the difference in the environment of O
and O1 atoms (OAlMg$_2$, O1Al$_2$Mg), one can trace the tendency of
shifting O$2s$ states with increasing amount of Al in the coordination
of O-atoms. The positions of the highest peaks in O$2s$ subbands for the
OAl$_3$, OAl$_2$Mg, OAlMg$_2$, OMg$_3$ coordinations are  $-1.818$,
$-1.45$, $-1.34$, $-1.29$ Ry ($-24.74$, $-19.73$, $-18.23$, $-17.55$
eV), respectively (see fig. \ref{f.bs}, \ref{f.pdos}, \ref{f.alp}). A
similar tendency can be determined also for O$2p$ states.

According to fig. \ref{f.alp}, the closest to Mg atoms N1, N, N2 and
Si3, Si4, Si2 (in the order of distances from Mg) are responsible for
the splitting of N$2s$ states into K and K' peaks, and their
contribution into K peak grows with approaching to Mg.

In spite of a small change in the \nef\ value for \Simgalon\ in
comparison with \Simgon, the composition of the TDOS at \ef\ differs
considerably. If in \Simgon\ the highest value of \nefi{i}\ belongs to
${\rm i}=\O$ atoms, in the \Simgalon\ SS it is not so high (16, 5
states/eV atom for O, O1, respectively; fig. \ref{f.alp}). The main
contribution to \nef\ is made by N6 (95) and N1 (54 states/eV atom). The
larger value of \nefi{N6} occurs because the Fermi energy is at a peak
for those atoms, whereas for N1 atoms it is on a slope. In the
\Simgalon\ SS, \nefi{i}\ tends to decrease with the distance from Mg
atoms (54 (N1), 14 (N2), 11 (N, N9), 8 (N8); 1.4 (Si3, Si4), 0.7 (Si2))
similarly to that in \Simgon. But this cannot take place for the atoms
closest to Al (95 (N6), 14 (N7), 16 (N5), 12 (N4), 19 (N3); 2.7 (Si,
Si1), 0.7 (Si5)). The data presented here reveal a localized character
of the free states. Therefore, we expect \Simgalon\ to be an insulator.

The cohesion energy of the \Simgalon\ SS is 4.17 eV/atom (table
\ref{t.oth}). This value can be used for the analysis of ordering in the
system in question. As the impurity channels in this system are at a
considerable distance from each other, we may assume that the
interaction between them is small. Hence, we admit that the cohesion
energy of the system of the composition \Simgalon, which includes
separate Al- and Mg-containing impurity channels, is expressed by the
formula $E^{Sep}_{coh} = 0.5\cdot(E_{coh}^{\Si_{6-x}\Al_x\O_x\N_{8-x}} +
E_{coh}^{\Si_{6-x}\Mg_x\O_x\N_{8-x}})$. Thus, the energy of ordering of
Al and Mg atoms is $E_o = E_{coh}^{Sep} -
E_{coh}^{\Si_{6-x}\Mg_{x/2}\Al_{x/2}\O_x\N_{8-x}} = 0.055$ eV/atom (3.15
eV/cell). This means that the ordering of Al and Mg atoms in the form of
alternating quasi-linear chains (fig. \ref{f.sol}a) is more favorable.

\section*{Conclusions}

The simulation of electronic and energy properties of \bSin, \Sialon\
and hypothetical ordered \Simgoon, \Simgons, \Simgon, \Simgalon\ SS allows
to make the following conclusions.

\begin{enumerate}

\item Comparison of the present results with the data of other
publications for \bSin\ and \Sialon\ systems exhibits a good quality
of the calculations performed.

\item The stability of the considered systems decrease in the following
order: \Sin\ $>$ \Sialon\ $>$ \Simgalon\ $>$ \Simgon\ $>$ \Simgoon\ $>$
\Simgons. The low stability of \Simgoon\ and \Simgons\ is in agreement
with experimental data and earlier theoretical studies.

\item The density of states at the Fermi energy in \Simgon\ is
$3.0\cdot 10^3$ states/eV cell, but it includes mainly localized
O$2p$, N$2p$ states, and the conductivity in \Simgon\ is unlikely.

\item Similarly to \Simgon, in the \Simgalon\ SS the number of states at \ef\
is considerable ($2.0\cdot 10^3$ states/eV cell), but they involve
mainly localized N$2p$ and to a less degree O$2p$ states, so, the
conductivity here is unlikely. It is established that the positions of
O$2s,p$ states in \Simgalon\ strongly depend on the coordination
environment of O atoms. Those states shift upward in the series
OAl$_3\rightarrow$OAl$_2$Mg$\rightarrow$OAlMg$_2\rightarrow$OMg$_3$. The
energy of ordering of Al and Mg atoms is estimated to be 3.15 eV/cell.

\item Considering various distributions of M and Mg atoms in
the SS of the general composition \Simgmon, we supposed the possibility
of the formation of quasi-linear, ring-like and spiral conductors, which
can be used, in particular, for quantum memory elements production.
However, the simulation of the \Simgalon\ SS shows that it does not
possess such properties. To find a particular composition of the
\Simgmon\ solid solution, we are going to study other possible
alternatives (${\rm M} = {\rm Sc, Ga, Y}$).

\end{enumerate}

\begin{acknowledgments}

This work was supported by the Russian Foundation for Basic Research,
grant \# 01-03-96515 (Ural).

\end{acknowledgments}

\end{document}